# Direct observation of internal quantum transitions and femtosecond radiative decay of excitons in monolayer WSe$_2$

C. Poellmann[1], P. Steinleitner[1], U. Leierseder[1], P. Nagler[1], G. Plechinger[1], M. Porer[1], R. Bratschitsch[2], C. Schüller[1], T. Korn[1] and R. Huber[1*]

[1] *Department of Physics, University of Regensburg, D-93040 Regensburg, Germany*

[2] *Institute of Physics, University of Münster, D-48149 Münster, Germany*

[*]*rupert.huber@ur.de*

**Atomically thin two-dimensional crystals have revolutionized materials science[1-3]. In particular, monolayer transition metal dichalcogenides promise novel optoelectronic applications, due to their direct energy gaps in the optical range[4-9]. Their electronic and optical properties, however, are complicated by exotic room-temperature excitons, whose fundamental structure and dynamics has been under intense investigation[10-18]. While interband spectroscopy probes energies of excitons with vanishing centre-of-mass momenta, the majority of excitons has remained elusive, raising questions about their unusual internal structure[13], symmetry[15-17], many-body effects[18], and dynamics. Here we report the first direct experimental access to all relevant excitons in single-layer $WSe_2$. Phase-locked mid-infrared pulses reveal the internal orbital 1s-2p resonance, which is highly sensitive to the shape of the excitonic envelope functions and provides accurate transition energies, oscillator strengths, densities and linewidths. Remarkably, the observed decay dynamics indicates a record fast radiative annihilation of small-momentum excitons within 150 fs, whereas Auger recombination prevails for optically dark states. The results provide a comprehensive view of excitons and introduce a new degree of freedom for quantum control, optoelectronics and valleytronics of dichalcogenide monolayers[19-25].**

Excitons emerge when photons promote electrons from the valence to the conduction band of a semi-conductor. Each hole left behind can bind with an electron by Coulomb attraction. Like atoms, excitons are described by their centre-of-mass momentum, $K$, and discrete quantum numbers characterizing the relative electron-hole motion (Fig. 1a). In inorganic bulk semiconductors, typical binding energies amount to few meV, due to the small effective mass and dielectric screening. In monolayer transition metal dichalcogenides (TMDC), such as $MoS_2$, $MoSe_2$, $WS_2$ or $WSe_2$, however, the two-dimensional electron confinement and suppressed screening leads to exotic, non-hydrogenic excitons with binding energies in excess of 0.2 eV [10-17]. Since these bound states remain stable at room temperature and dominate many optical and electronic properties of monolayer TMDCs, a detailed microscopic understanding is indispensable for almost all prospective applications of TMDCs.

Spectroscopy with visible light probing electronic transitions across the direct energy gap, $E_g$, has provided key insight into the energy levels of bound states[2,11,13,14]. For instance, radiative recombination of bound electron-hole pairs causes photoluminescence (PL)[4-7] at a photon energy that is reduced with respect to $E_g$, by the exciton binding energy. Furthermore, the small exciton Bohr radius leads to anomalously strong interband absorption[11-13,16,17,26]. Even excited and charged exciton states, called trions, are imprinted on interband spectra[2,10,14,15,22,27]. Yet, the negligible photon momentum restricts optical transitions to creating or annihilating excitons with $K \approx 0$ (Fig. 1a, blue vertical arrow). Due to scattering, the lion's share of excitons can assume finite momenta, making them optically dark. Additionally, interband excitation probes the envelope functions of bound electron-hole pairs only at vanishing electron-hole distance (Supplementary Equation 1) and evaluating interband dipole moments requires sophisticated band structure calculations[14]. Extracting quantitative information about densities, internal structure, many-body interactions, and ultrafast dynamics has, therefore, remained challenging.

Here we investigate a new class of quantum excitations in WSe$_2$: Internal transitions connecting different orbital states of bound electron-hole pairs[28,29] (see Fig. 1a, red arrows) by absorption of mid-infrared photons. This approach represents the first direct, destruction-free access to all excitons created in a monolayer, irrespective of $K$. The mid-infrared absorption coefficient depends sensitively on the full spatial profile of the exciton's envelope function, but remains independent of interband dipole moments (Supplementary Equation 2). It allows us to establish a detailed picture of the internal structure of excitons in a monolayer TMDC, including absolute values for 1s-2p transition energies, oscillator strengths, particle densities, many-body interactions and ultrafast dynamics. Importantly, the giant interband dipole moment[26] leads to an extremely efficient radiative decay of excitons at $K \approx 0$.

Our WSe$_2$ samples are manufactured by mechanical exfoliation and transfer onto a CVD diamond window using a recently developed deterministic process[30]. Single-layer flakes with lateral dimensions above 70 μm are identified by optical microscopy (Fig. 1b) and PL mapping (Fig. 1c). The corresponding room-temperature PL spectrum (Fig. 1d, blue solid curve) features a peak wavelength of 750 nm, a width of 20 nm (FWHM) and a Stokes shift of 6 nm with respect to the absorption maximum (Supplementary Fig. 1), characteristic of the 1s state of the A exciton[7]. To explore the dynamics of these quasiparticles, we employ a 90-fs laser pulse centred at $\lambda_c$ = 742 nm (Fig. 1d, blue dashed curve) to selectively inject excitons at $K$ = 0. After a variable delay time $t_{PP}$, the low-energy dielectric response is probed by a phase-locked mid-infrared pulse (Fig. 1a), whose complete waveform is mapped out electro-optically[29] (see methods) as a function of the recording time, $t_{EOS}$. Figure 1e displays the mid-infrared field transient transmitted through the unexcited sample, $E_{ref}$ (black curve), and its pump-induced change, $\Delta E$ (red curve) recorded at a fixed delay time $t_{PP}$ = 75 fs. $\Delta E$ roughly traces the reference pulse with a phase offset of $\pi$, indicating a pump-induced reduction of the amplitude of the transmitted probe field.

These transients mark the first observation of the mid-infrared exciton response in a single-layer crystal and allow us to retrieve the full dielectric response of the excited sample[29], characterized by the real parts of the optical conductivity $\Delta\sigma_1$ (Fig. 2a) and the dielectric function $\Delta\varepsilon_1$ (Fig. 2b). $\Delta\sigma_1$, which is proportional to the mid-infrared absorption coefficient, exhibits a broad maximum while $\Delta\varepsilon_1$ follows a dispersive shape with a zero crossing at a photon energy of 165 meV. These spectra are in stark contrast to the Drude-like response of free electron-hole pairs observed in bulk $WSe_2$ (Supplementary Fig. 2) and prove the existence of a new mid-infrared quantum transition in monolayer TMDCs – the 1s-2p orbital exciton resonance, as seen below. Figures 2c-f show the pump-induced infrared response at $t_{PP}$ = 275 fs and 1.6 ps, respectively. While the qualitative resonance features are similar to Figs. 2a and b, the amplitude of $\Delta\sigma_1$ (Figs. 2c,d) and the slope of $\Delta\varepsilon_1$ (Figs. 2d,f) is reduced and the resonance slightly narrows and blue shifts with increasing $t_{PP}$.

For a quantitative analysis we describe the Wannier excitons by a two-dimensional hydrogen model (Supplementary Information). Nonlocal dielectric screening capturing the size-dependent extent of the electric field between electron and hole into the dielectric environment renders the interaction potential non-Coulombic[13,15] (Fig. 2h). This modification enhances the 1s Bohr radius from 4 Å to 8.3 Å, whereas the 2p wavefunction is less affected (Fig. 2g). For single-layer $WSe_2$, we compute an energy separation between the 1s and 2p states of 170 meV (Fig. 2h). The oscillator strength $f_{1s-2p}$ = 0.27 of the internal resonance of a single electron-hole pair is obtained by spatial convolution of the dipole operator with the full excitonic envelope functions (Supplementary Equation 2). Neglecting nonlocal screening would reduce $f_{1s-2p}$ by a factor of 5. To match the experimental data of Fig. 2, we introduce a Lorentzian broadening of the 1s-2p oscillator (Supplementary Information) and use the 1s exciton density $n_X$, the resonance energy $E_{res}$, and the linewidth $\Delta$ as adjustable parameters fitting simultaneously $\Delta\sigma_1$ and $\Delta\varepsilon_1$ (shaded curves in Figs. 2a-f).

For $t_{PP}$ = 75 fs and 275 fs (Figs. 2a-d), the model perfectly describes both response functions assuming $E_{res}$ = 162 meV and 179 meV, respectively. Moreover, the 1s exciton density $n_X$ obtained at $t_{PP}$ = 75 fs matches quantitatively with the density of absorbed pump photons of $n_a = (3.0 \pm 0.4) \times 10^{12}$ cm$^{-2}$ (Supplementary Information). These observations confirm that (i) our assignment of the infrared response to the 1s-2p transition is correct, (ii) the optical pump exclusively creates 1s excitons, and (iii) all 1s excitons are fully accounted for by the infrared response. Interestingly, the model fit cannot reproduce an additional fine structure discernible at $t_{PP}$ = 1.6 ps (Figs. 2e,f), in the energy window between 125 and 165 meV. We suggest that these features may be caused by the formation of trions or trapped excitons, where additional electronic resonances are expected, in analogy to the hydrogen anion[2,10,14,15,22,27].

Figure 3 summarizes the fitting parameters extracted from response functions at different pump fluences $\Phi$ and fixed $t_{PP}$ = 75 fs (Supplementary Fig. 3). $n_X$ (Fig. 3a, red spheres) agrees perfectly with the density of absorbed pump photons (broken line) and shows no sign of saturation. Remarkably, a slight red-shift of $E_{res}$ (Fig. 3b) and a strong increase of the linewidth $\Delta$ from 120 to 180 meV (Fig. 3c) is seen with increasing density, even though the maximum value of $n_X = 7.3 \times 10^{12}$ cm$^{-2}$ is still one order of magnitude below the Mott density estimated by close packing of spheres. The broadening of the 1s-2p transition is likely dominated by the width of the spatially more extended 2p state. This assumption is supported by the relatively narrow PL linewidth of 45 meV.

To track the dynamics of the density $n_X$ as a function of $t_{PP}$ more systematically, we record the maximum pump-induced change of the probe field, $\Delta E_{max}$, at fixed $t_{EOS}$ = 12 fs (Fig. 1e, red arrow), which is proportional to $n_X$ (see discussion of Supplementary Fig. 4), and scan $t_{PP}$. For all pump fluences the exciton density sets on rapidly within the duration of the pump pulse of 90 fs (Fig. 4a) and decays subsequently in two distinct steps: A sub-ps dynamics (blue shaded region, Fig. 4a) is followed by a non-exponential decay dominating for $t_{PP}$ > 0.3 ps (Fig. 4b). The latter is well described by the bi-molecular rate equation $\partial n_X/\partial t = -(\gamma/2) n_X^2$, indicating non-radiative Auger recombination[18]. Its efficiency is strongly enhanced by the reduced dimensionality and weak screening. The bi-molecular decay is faithfully fitted (Fig. 4b, red solid curves) using the same universal decay constant $\gamma = (0.13 \pm 0.1)$ cm$^2$/s for all pump fluences. This value is comparable with results obtained from interband spectroscopy of other TMDCs[18].

Remarkably, subtracting the bi-molecular decay (red solid curves) leaves a subset of excitons (Fig. 4a, blue shaded areas) that decays exponentially (black dashed curves) with an extremely short time constant of $\tau = (150 \pm 20)$ fs. Ultrafast trapping or scattering of excitons cannot cause this dynamics since it does not change $n_X$. Also hypothetical ultrafast non-radiative recombination of excitons can be ruled out as a microscopic origin, since such a process would affect all excitons and continue until they are completely annihilated. It would not explain the observed decomposition into two subsets decaying with different dynamics. Since we observe the same relative weight of fast and slow dynamics at all spots of all WSe$_2$ samples tested, spatial inhomogeneity plays no role either. In contrast, a strong inherent inhomogeneity has been predicted in momentum space[26]: Excitons with $K \approx 0$ can couple efficiently with light, whereas large-momentum states are optically dark. The interband dipole moment has been suggested to be sufficiently large for radiative recombination to become faster than any non-radiative decay. Our experimental data are characteristic of this unique novel scenario and the time constant $\tau$ found above is similar to radiative decay times computed for MoS$_2$[26].

The ultrashort radiative decay should broaden optically bright 1s states by 9 meV, contributing to the total PL linewidth. The homogeneous broadening should also add to the 1s-2p linewidth $\Delta$, yet only for $K \approx 0$. Although $\Delta$ is always dominated by the width of the 2p state, the 1s-2p resonance of a bright exciton ensemble at $t_{PP}$ = 75 fs (Fig. 3c, red spheres) is indeed tentatively broader than the corresponding resonance at $t_{PP}$ = 0.8 ps and 1.6 ps, when the excitons are mostly dark (Fig. 3c, black triangles).

Once excitons are optically created at $K = 0$ they may either recombine radiatively (Fig. 4c) or scatter with phonons into dark states outside the light cone. Large-momentum states are still visible to the probe pulse, but their radiative annihilation is blocked and the fastest remaining decay channel is given by Auger recombination (Fig. 4d). A complete rate equation model (Supplementary Fig. 5) allows us to determine the momentum scattering rate converting bright into dark particles as $\Gamma = 3.7$ ps$^{-1}$. This value is consistent with typical exciton mobilities (Supplementary Information). Note that the ratios between the scattering rate $\Gamma$, the Auger rate $\gamma$ and the radiative recombination rate $\tau^{-1}$ define the ultimate quantum yield achievable in photonic and optoelectronic TMDC devices exploiting radiative interband transitions.

In conclusion, the first direct observation of Rydberg-like internal exciton absorption in a single-layer TMDC allows us to trace both optically dark and bright states. The data provide absolute values of internal transition energies, oscillator strengths, and densities and quantify many-body effects. Most remarkably, we reveal the fingerprint of a uniquely fast radiative annihilation of 1s excitons with vanishing momentum and efficient Auger recombination of dark excitons. The extremely strong coupling of excitons to light opens exciting perspectives for quantum electrodynamics applications[9]. Furthermore, the internal orbital resonances offer a new handle for sophisticated quantum control of monolayer TMDCs.

**Methods**

The investigated system is a monolayer of $WSe_2$ on a CVD diamond window. Manufactured by mechanical exfoliation and subsequent deterministic transfer[30] from a viscoelastic substrate onto diamond, typical monolayer flakes feature sizes of up to 300 μm × 70 μm. For our experiments, 12-fs light pulses with a centre wavelength of 800 nm are derived from a high-repetition-rate Ti:sapphire amplifier system. A first part of the laser output is filtered by a bandpass with a centre wavelength of 742 nm and a bandwidth of 9 nm resulting in 90-fs pulses, which selectively inject 1s A excitons in the $WSe_2$ monolayer. Another part of the laser pulses generates few-cycle mid-infrared probe transients via optical rectification in a 50 μm thick $AgGaS_2$ emitter covering a photon energy range from 125 to 210 meV. The probe pulses are focussed onto the sample. The diffraction-limited mid-infrared focal spot (FWHM diameter < 35 μm) is set to be distinctly smaller than the area of the monolayer flakes and the optical pump spot, to probe a homogeneously excited sample area. Extending the concept of electro-optic sampling to the mid-infrared spectral range allows us to trace the absolute amplitude and phase of the probe transient, transmitted through the excited and unexcited sample. $\Delta\sigma_1$ and $\Delta\varepsilon_1$ can be extracted independently from each other by a transfer matrix formalism, without resorting to a Kramers-Kronig analysis. In case of a monolayer this analysis is particularly precise since typical etalon artefacts inside the sample are absent.

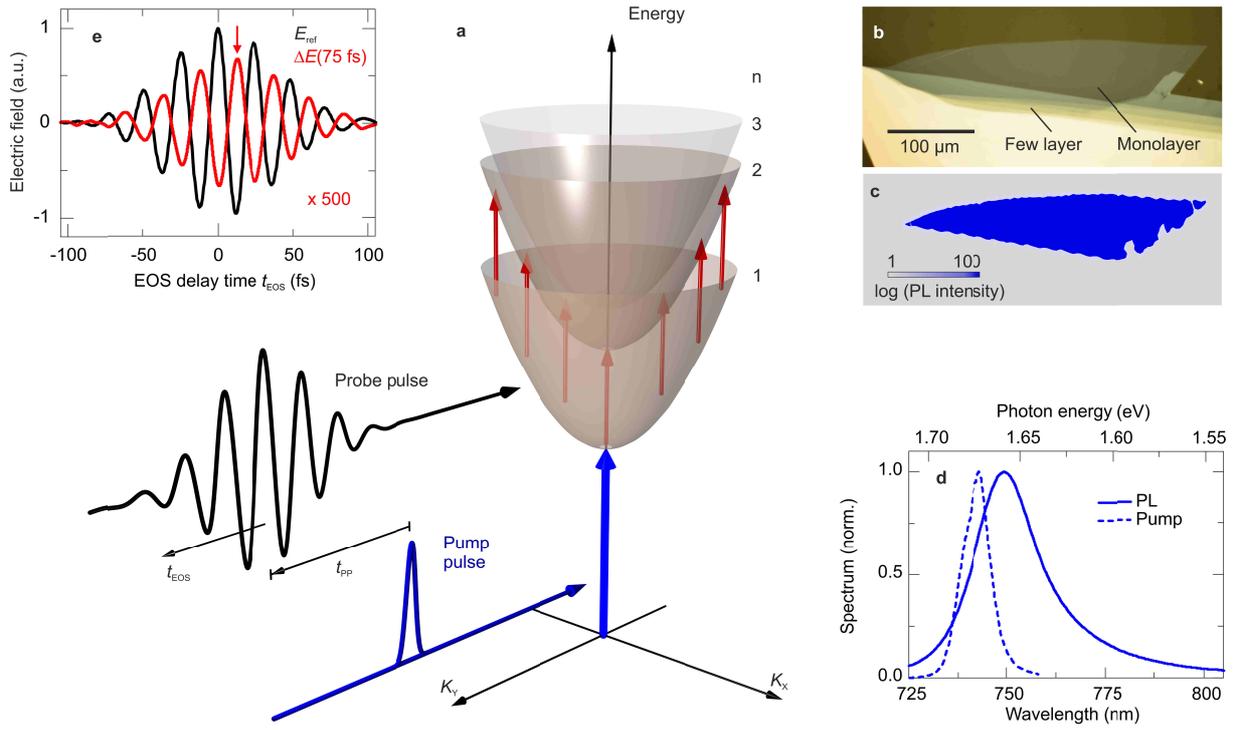

**Figure 1 | Intra- and interband spectroscopy of single-layer WSe$_2$. a**, Schematic dispersion (amber paraboloids) of excitons with different principal quantum numbers n, as a function of the centre-of-mass momentum $K = \sqrt{K_X^2 + K_Y^2}$. Resonant interband photon absorption (blue arrow) creates 1s excitons at $K = 0$, whereas mid-infrared absorption probes the internal 1s-2p transition (red arrows) of all pre-existing excitons, irrespective of $K$. In the time resolved pump-probe experiment, 1s A excitons are resonantly generated by a 90-fs near-infrared pump pulse (blue) while a mid-infrared transient (black waveform) delayed by $t_{PP}$ probes their internal response. The oscillating carrier waveform of the probe pulse is directly mapped out as a function of the delay time $t_{EOS}$, by ultrabroadband electro-optic sampling. **b, c**, Optical microscopy image (**b**) and PL intensity map (**c**) of an exfoliated WSe$_2$ monolayer on a viscoelastic substrate before the transfer onto the diamond window (**b**) and thereafter (**c**). **d**, Measured photoluminescence spectrum (blue solid line) of the monolayer sample excited by a continuous wave laser at a wavelength of 532 nm. Dashed blue line: Spectrum of the femtosecond pulse used in the pump-probe experiment. **e**, Waveform of the probe pulse $E_{ref}$ (black curve) transmitted through the unexcited WSe$_2$ monolayer and pump-induced change $\Delta E$ (red curve, scaled up by a factor of 500) at $t_{PP} = 75$ fs after resonant injection of 1s excitons. The red arrow indicates the maximum of $\Delta E(t_{EOS})$. All experiments are performed at room temperature.

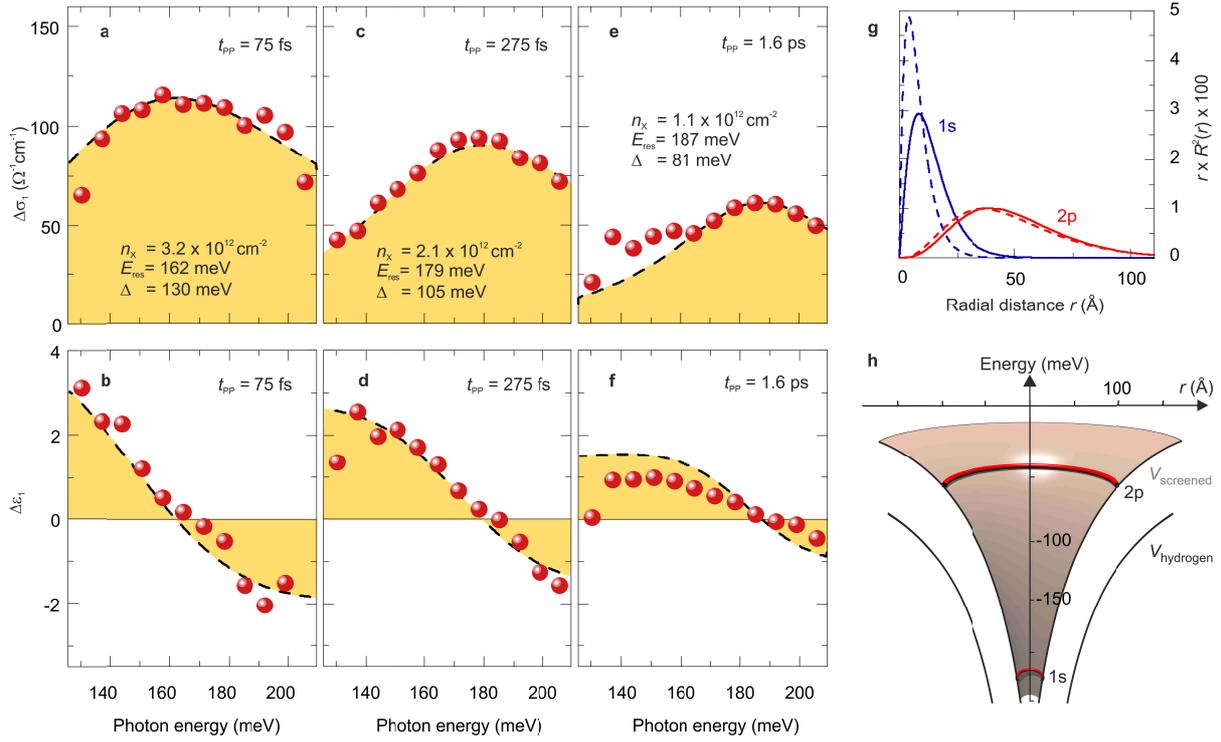

**Figure 2 | Time-resolved response of the intra-excitonic 1s-2p transition. a**, **c**, **e**, Real part of the pump-induced mid-infrared conductivity $\Delta\sigma_1(\hbar\omega)$ of photoexcited monolayer WSe$_2$, as a function of the photon energy, $\hbar\omega$, for three different delay times $t_{PP}$ (pump fluence, $\Phi = 16$ µJ/cm²). **b**, **d**, **f**, Corresponding real part of the dielectric function, $\Delta\varepsilon_1(\hbar\omega)$. Red spheres: Experimental data; black dashed curve: Two-dimensional Wannier exciton model simultaneously fitting $\Delta\sigma_1$ and $\Delta\varepsilon_1$. The adapted parameters of 1s exciton density $n_X$, transition energy $E_{res}$, and linewidth $\Delta$, are given in panels **a**, **c**, and **e**. **g**, Radial electron distribution of the 1s (blue) and the 2p (red) exciton for a two-dimensional model assuming a locally screened (dashed) and a nonlocally screened (solid) potential, as a function of the radial distance $r$. **h**, Interaction potential $V_{screened}$ taking into account nonlocal Coulomb screening in a WSe$_2$ monolayer (light brown hypersurface). The calculated eigenenergies of the 1s and 2p exciton states of -225 meV and -55 meV, respectively, are indicated by red lines. $V_{hydrogen}$ indicates the bare 2D Coulomb potential.

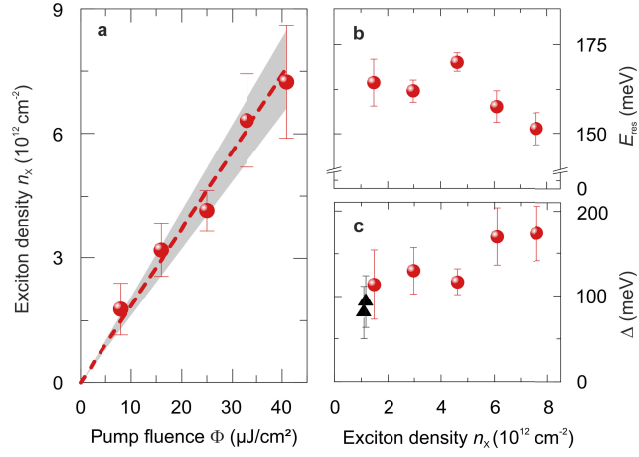

**Figure 3 | Density-dependent renormalization of the internal excitonic response. a**, 1s-exciton density $n_X$, as a function of the pump fluence $\Phi$. Red spheres: Values of $n_X$ as extracted by fitting the mid-infrared response functions measured for pump delay time $t_{PP}$ = 75 fs (see Figs. 2a, b and Supplementary Fig. 3) with the two-dimensional Wannier model described in the text. Dashed line: Values of $n_X$ as obtained from the pump fluence and the measured absorptivity (Supplementary Information). Grey area: Experimental uncertainty. **b, c**, Resonance frequency $E_{res}$ (**b**) and linewidth $\Delta$ (**c**) as a function of density $n_X$. Red spheres: Data extracted by fitting the mid-infrared response functions for $t_{PP}$ = 75 fs. Black triangles in **c**: Linewidths obtained for delay times $t_{PP}$ = 0.8 ps and 1.6 ps.

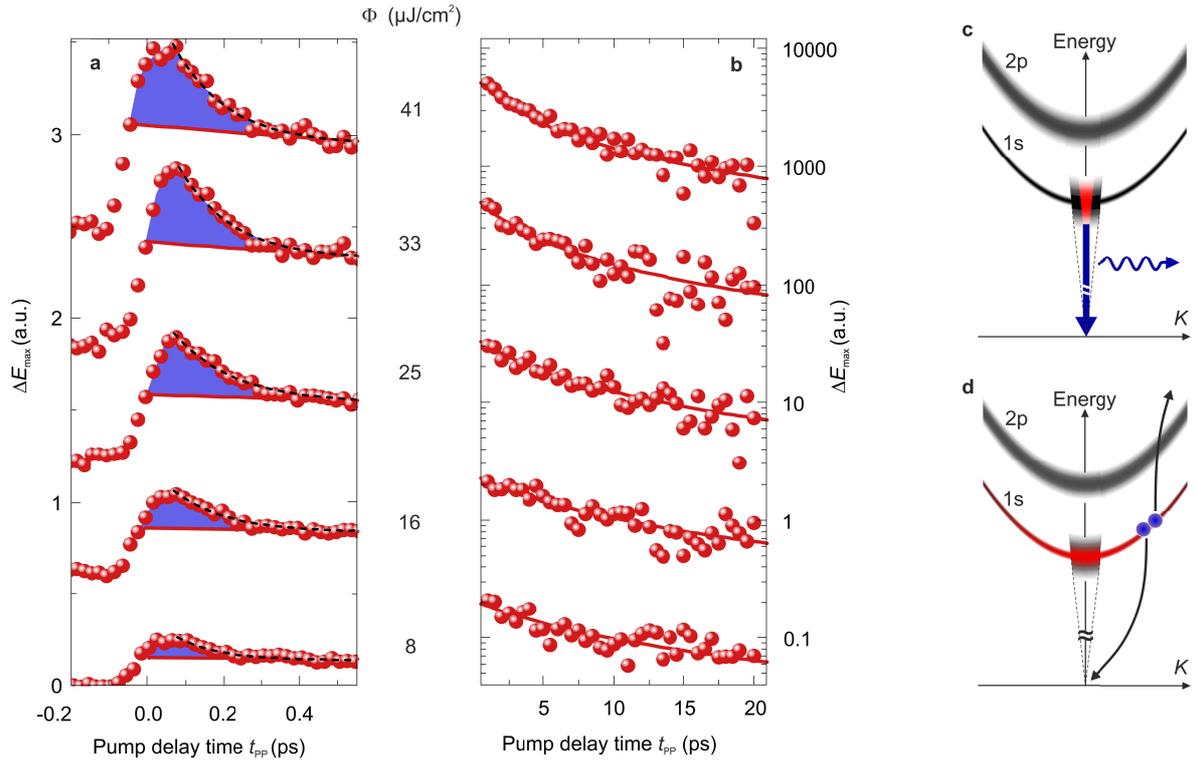

**Figure 4 | Ultrafast dynamics of exciton density in single-layer WSe$_2$. a**, The maximum of the pump-induced change $\Delta E_{max}$ recorded at a fixed electro-optic sampling time $t_{EOS}$ = 12 fs is proportional to the density $n_X$ of 1s excitons (Supplementary Information). $\Delta E_{max}$ is displayed as function of the pump delay time $t_{PP}$ < 0.55 ps for different fluences Φ indicated on the right. Curves corresponding to different Φ are vertically offset for better visibility. Red spheres: Experimental data, red solid curves: Bi-molecular decay model (γ = 0.13 cm²/s), blue shaded area: Subset of excitons following an exponential decay (τ = 150 fs). **b**, Semilogarithmic representation of $\Delta E_{max}$ as in panel **a**, for delay times $t_{PP}$ > 0.55 ps. **c**, Schematic of the ultrafast radiative interband recombination (blue arrow) of optically bright 1s excitons (red area) at $K \approx 0$ within the light cone (black dashed lines). **d**, Non-radiative exciton annihilation by Auger recombination (black arrows) of a dark population (red area) of excitons (blue spheres with halo) with large centre-of-mass momenta.

**Acknowledgements**

We thank M. Eisele, T. Cocker, M. Huber, J. Fabian, A. Chernikov, S. Michaelis and R. Schmidt for helpful discussions and M. Furthmeier for technical assistance. Support by the European Research Council through ERC grant 305003 (QUANTUMsubCYCLE) and by Deutsche Forschungsgemeinschaft (DFG) through Graduate Research College GK1570 and KO3612/1-1 is acknowledged.

**Author contributions**

C.P., R.B., C.S., T.K. and R.H. planned the project; P.N., G.P., R.B., C.S. and T.K. provided, processed and characterised the samples; C.P., P.S., U.L. and M.P. performed the femtosecond measurements; C.P., P.S., U.L., M.P. and R.H. analysed the data; C.P., P.S. and R.H. elaborated the theoretical model; C.P., P.S. and R.H. wrote the paper with contributions from all authors.

**Competing financial interests**

The authors declare no competing financial interests.